\documentclass[aps,prd,twocolumn,superscriptaddress,nofootinbib,floatfix,10pt]{revtex4-2}

\usepackage{amsmath,amssymb}
\usepackage{bm}
\usepackage{graphicx}
\usepackage{xcolor}
\usepackage{booktabs}
\usepackage[colorlinks=true,citecolor=blue,linkcolor=blue,urlcolor=blue]{hyperref}

\begin{document}

\title{Migdal Ionization as a Probe of Light Dark Matter from Nuclear Transition}

\author{Yuanchao Lou}
\email{yuanchao_lou@nnu.edu.cn}
\affiliation{Department of Physics and Institute of Theoretical Physics, Nanjing Normal University, Nanjing, 210023, China}

\author{Liangliang Su}
\email{liangliang.su@kit.edu}
\affiliation{Institute for Astroparticle Physics, Karlsruhe Institute of Technology, D-76131 Karlsruhe, Germany}

\author{Lei Wu}
\email{leiwu@njnu.edu.cn}
\affiliation{Department of Physics and Institute of Theoretical Physics, Nanjing Normal University, Nanjing, 210023, China}
\affiliation{Nanjing Key Laboratory of Particle Physics and Astrophysics, Nanjing, 210023, China}

\begin{abstract}

Nuclear reactors serve as a key artificial source of light dark matter. Direct detection of reactor-produced dark matter faces substantial obstacles, since quenching effects suppress conventional elastic scattering signals below detector thresholds. We present a new search strategy utilizing the Migdal effect in germanium detectors to probe light dark matter produced via nuclear de-excitation from reactors. Using ON–OFF residual spectra from the TEXONO experiment, we set a new stringent limit on the dark matter and nucleus interaction  over the mass range $0.01\,\text{MeV}\le m_\chi
\lesssim2.6\,\text{MeV}$, which provides a complementary bound to existing cosmological and astrophysical limits.

\end{abstract}

\maketitle

\section{Introduction}
\label{sec:intro}

The existence of dark matter (DM) is firmly established by overwhelming gravitational evidence on astrophysical and cosmological scales. However, its fundamental nature at the microscopic level remains one of the most profound mysteries in physics. For decades, the search has been dominated by the Weakly Interacting Massive Particle (WIMP) paradigm~(see recent reviews, e.g.~\cite{Bertone:2004pz,Battaglieri:2017aum}). Yet, the continued absence of a confirmed signal in conventional WIMP searches~\cite{LUX:2016ggv,
XENON:2018voc,PandaX-4T:2021bab} has motivated a significant shift in the field towards exploring lighter DM candidates~\cite{Boehm:2003hm,Pospelov:2007mp,Arkani-Hamed:2008hhe,
Holdom:1985ag,Alexander:2016aln,Hochberg:2014dra,Hochberg:2014kqa,Kuflik:2015isi,
Kuflik:2017iqs,Griest:1990kh,DAgnolo:2015ujb,Dror:2016rxc,
DAgnolo:2017dbv}. This poses a significant challenge for direct detection, as the resulting nuclear recoils fall far below current detector thresholds, necessitating novel detection strategies (see examples, e.g.~\cite{Essig:2012yx,Essig:2015cda,An:2017ojc,Knapen:2017xzo,Bringmann:2018cvk,Cappiello:2019qsw,Alvey:2019zaa,Su:2020zny,
Ge:2020yuf,Kahn:2021ttr,Wang:2021nbf,Su:2023zgr,
Nagao:2024hit,Liang:2024xcx}).

The Migdal effect provides a viable mechanism to circumvent this limitation. During a DM-nucleus scattering event, the sudden acceleration of the nucleus can lead to the ionization or excitation of atomic electrons~\cite{Migdal:1939,Migdal:1941}. This process produces an observable electronic signal, with energy in the eV–keV range, in addition to the tiny nuclear recoil. Later works has revealed its importance for the dark matter direct detection~\cite{Vergados:2005dpd,Moustakidis:2005gx,Ejiri:2005aj,Bernabei:2007jz,Ibe:2017yqa,Dolan:2017xbu,Bell:2019egg,Essig:2019xkx,Baxter:2019pnz,Liang:2019nnx,Flambaum:2020xxo,Liu:2020pat,Knapen:2020aky,Liang:2020ryg,Liao:2021yog,Acevedo:2021kly,Wang:2021oha,Li:2022acp,Cox:2022ekg,Liang:2022xbu,Blanco:2022pkt,Tomar:2022ofh,Adams:2022zvg,Berghaus:2022pbu,Gu:2023pfg,Xu:2023wev,Qiao:2023pbw,Li:2023xkf,Herrera:2023xun,Dzuba:2023zcq,Kahn:2024nyv,Kang:2024kec,Nakano:2024oon,Maity:2024hzb,Esposito:2025iry,Mai:2025zau,Berghaus:2026kmj,Kahlhoefer:2026usv,NellenMondragon:2026thr}. Leveraging the Migdal effect, several searches for sub-GeV dark matter have been performed~\cite{LUX:2018akb,CDEX:2019hzn,XENON:2019zpr,COSINE-100:2021poy,DarkSide:2022dhx,EDELWEISS:2022ktt,SuperCDMS:2023sql,PandaX:2023xgl}. However, Migdal effect itself remained experimentally unverified in neutron-nucleus scattering until recently, when it was observed in a gaseous detector~\cite{Yi:2026fmf}. This breakthrough has spurred further validation in liquid xenon experiments~\cite{Xu:2026acq} and other detectors~\cite{MIGDAL:2022yip,Nakamura:2020kex}.

While direct detection experiments have made significant progress in searching for light dark matter in the galactic halo, nuclear reactors provide a complementary and unique source characterized by an intense high flux of light new particles~\cite{lou2026probingdarkphotonsnuclear,
gao2025constraintsmillichargedparticlesnuclear,NEON:2024bpw,Ge:2017mcq,Dai:2025kai,Gong:2026dte}. Unlike the non-relativistic halo dark matter, the dark matter particles produced in nuclear reactors, arising via the decay of light mediators emitted during nuclear de-excitation, $N^*\to N+V(\to \chi\bar{\chi})$, can possess MeV-scale energies since the mediator energy is
fixed by the discrete nuclear transition energy, $E_V\simeq\omega_i$. This high-flux, relativistic component offers a distinct opportunity to probe the sub-MeV parameter space that is inaccessible to standard halo searches. However, detecting these reactor-produced particles via standard elastic scattering in semiconductor detectors is hindered by the quenching effect. Since a substantial fraction of the nuclear‑recoil energy is dissipated into lattice vibrations, the deposited ionization energy in electron equivalent is typically only a few hundred ${\rm eV_{ee}}$. This lies below the threshold of conventional experiments, such as TEXONO germanium detector~\cite{TEXONO:2018nir}.

In this work, we study Migdal ionization as an observable channel for
reactor-produced sub-MeV dark matter in low-threshold germanium
detectors. The sudden acceleration of the nucleus can tear off bound electrons, generating an accompanying ionization signal that raises the deposited energy above threshold. The incoming dark matter flux is computed from neutron-capture nuclear de-excitation in $^{238}{\rm U}$ and $^{10}{\rm
B}$. Using the TEXONO reactor ON--OFF residual
spectrum~\cite{TEXONO:2018nir}, we derive 95\% C.L.\ limits on the
reference dark matter--proton cross section $\bar{\sigma}_{\chi p}$
for dark matter masses $0.01\,{\rm MeV}<m_\chi<2.6\,{\rm MeV}$. Importantly, since the incident \(\chi\) flux is generated in the
reactor, the limits derived here are independent of the cosmological
abundance of \(\chi\) and do not require \(\chi\) to constitute all the observed dark matter density. This paper is organized as
follows. Section~\ref{sec:production} describes reactor nuclear de-excitation
and the resulting dark matter flux. Section~\ref{sec:detection}
presents the Migdal signal and the resulting TEXONO limits. We
conclude in Sec.~\ref{sec:conclusion}.

\section{Dark Matter Production}
\label{sec:production}

We consider a kinetically mixed dark photon $V$~\cite{Holdom:1985ag}
coupled to a Dirac fermion dark matter state $\chi$. The relevant
Lagrangian is
\begin{equation}
\mathcal{L}
\supset
-\frac{1}{4}V_{\mu\nu}V^{\mu\nu}
+\frac{1}{2}m_V^2V_\mu V^\mu
-\frac{\epsilon}{2}V_{\mu\nu}F^{\mu\nu}
+g_\chi V_\mu\bar{\chi}\gamma^\mu\chi,
\label{eq:lagrangian}
\end{equation}
where $g_\chi=\sqrt{4\pi\alpha_D}$, $m_V$ is the dark photon mass, which
can be generated via the Stueckelberg or dark Higgs
mechanism~\cite{Kors:2004dx,Feldman:2007wj,Morrissey:2009ur}, and
$\epsilon$ is the kinetic mixing parameter. After diagonalizing the
gauge kinetic terms, $V$ couples to the electromagnetic current with
strength $\epsilon e$. We focus on the invisible decay regime
$m_V>2m_\chi$ and take $\alpha_D\gg\epsilon^2\alpha$, so that
${\rm BR}(V\to\chi\bar{\chi})\simeq1$. For the parameter space considered here, the mediator decays promptly
near the reactor source.

In the vector-portal scenario, neutron-capture de-excitation provides
a line source of vector mediators: an excited nuclear state $N^*$ can
de-excite by emitting either a photon or, through kinetic mixing, an
on-shell vector mediator $V$, via $n+N\to N^*\to N+V$, followed by
$V\to\chi\bar{\chi}$. The emitted mediator carries energy
$E_{V,i}\simeq\omega_i$, where $\omega_i$ is the $i$-th nuclear
transition energy, with nuclear recoil corrections negligible at MeV
scales. On-shell production is kinematically allowed whenever
$m_V<\omega_i$.

For the $i$-th transition, the mediator production rate at the reactor
is
\begin{equation}
\dot{N}_{V,i} = \dot{N}_{\gamma,i}\,r_{E1,i},
\label{eq:NV}
\end{equation}
where $\dot{N}_{\gamma,i}$ is the ordinary photon emission rate and
$r_{E1,i}$ is the dark-vector-to-photon emission ratio. For an E1
transition of energy $\omega_i$~\cite{Pitrou:2019pqh,
gao2025constraintsmillichargedparticlesnuclear,
lou2026probingdarkphotonsnuclear},
\begin{equation}
r_{E1,i}
=
\epsilon^2
\left(1+\frac{m_V^2}{2\omega_i^2}\right)
\sqrt{1-\frac{m_V^2}{\omega_i^2}}\,
\Theta(\omega_i-m_V).
\label{eq:rE1}
\end{equation}
The ordinary photon emission rate is
\begin{equation}
\dot{N}_{\gamma,i}
=
N_n\,
\frac{Y_n^{(N)}}{\sum_N Y_n^{(N)}}\,
\frac{I_\gamma^{(N,\omega_i)}}{100}.
\label{eq:Ngamma}
\end{equation}
Here $N_n = 7.8\times10^{19}\,P({\rm GW})~{\rm s}^{-1}$ is the total
neutron production rate, corresponding to an average fission energy of
$200~{\rm MeV}$ and 2.5 neutrons per
fission~\cite{schunck2024nuclear,TEXONO:2005fmk}. For the TEXONO
benchmark with $P=2.9~{\rm GW}$, $N_n\simeq2.3\times10^{20}~{\rm
s}^{-1}$. $Y_n^{(N)}$ is the neutron-capture yield per fission for isotope $N$,
and $I_\gamma^{(N,\omega_i)}$ is the number of photons of energy
$\omega_i$ emitted per 100 radiative neutron captures on that
isotope~\cite{TEXONO:2005fmk,ENSDF,Mughabghab:2006}. For $^{10}{\rm
B}$, this  photon yield includes the small radiative-capture
contribution to $^{10}{\rm B}(n,\gamma){}^{11}{\rm B}$, while the
dominant $^{10}{\rm B}(n,\alpha){}^{7}{\rm Li}$ channel does not
contribute to the MeV-scale gamma lines 
here~\cite{Mughabghab:2006}.

The two-body decay $V\to\chi\bar{\chi}$ produces a box-shaped dark
matter spectrum with endpoints
\begin{equation}
E_{\chi,i}^{\rm min,max}
=
\frac{E_{V,i}}{2}
\mp
\frac{\sqrt{E_{V,i}^2-m_V^2}\sqrt{m_V^2-4m_\chi^2}}{2m_V},
\label{eq:Echi_minmax}
\end{equation}
and width $\Delta E_{\chi,i}=E_{\chi,i}^{\rm max}-E_{\chi,i}^{\rm
min}$. The differential dark matter flux at the detector from the
$i$-th transition is
\begin{equation}
\frac{d\Phi_{\chi,i}}{dE_\chi}
=
\frac{2\,\dot{N}_{V,i}\,{\rm BR}(V\to\chi\bar{\chi})}
{4\pi L^2\,\Delta E_{\chi,i}},
\quad
E_{\chi,i}^{\rm min}\leq E_\chi\leq E_{\chi,i}^{\rm max},
\label{eq:flux_i}
\end{equation}
where \(L=28~{\rm m}\) is the baseline between the Kuo-Sheng reactor core and the TEXONO germanium detector~\cite{TEXONO:2018nir},the factor
of two accounts for both $\chi$ and $\bar{\chi}$, and the total flux
is $d\Phi_\chi/dE_\chi=\sum_i d\Phi_{\chi,i}/dE_\chi$.

\begin{figure}[t]
\centering
\includegraphics[width=\columnwidth]{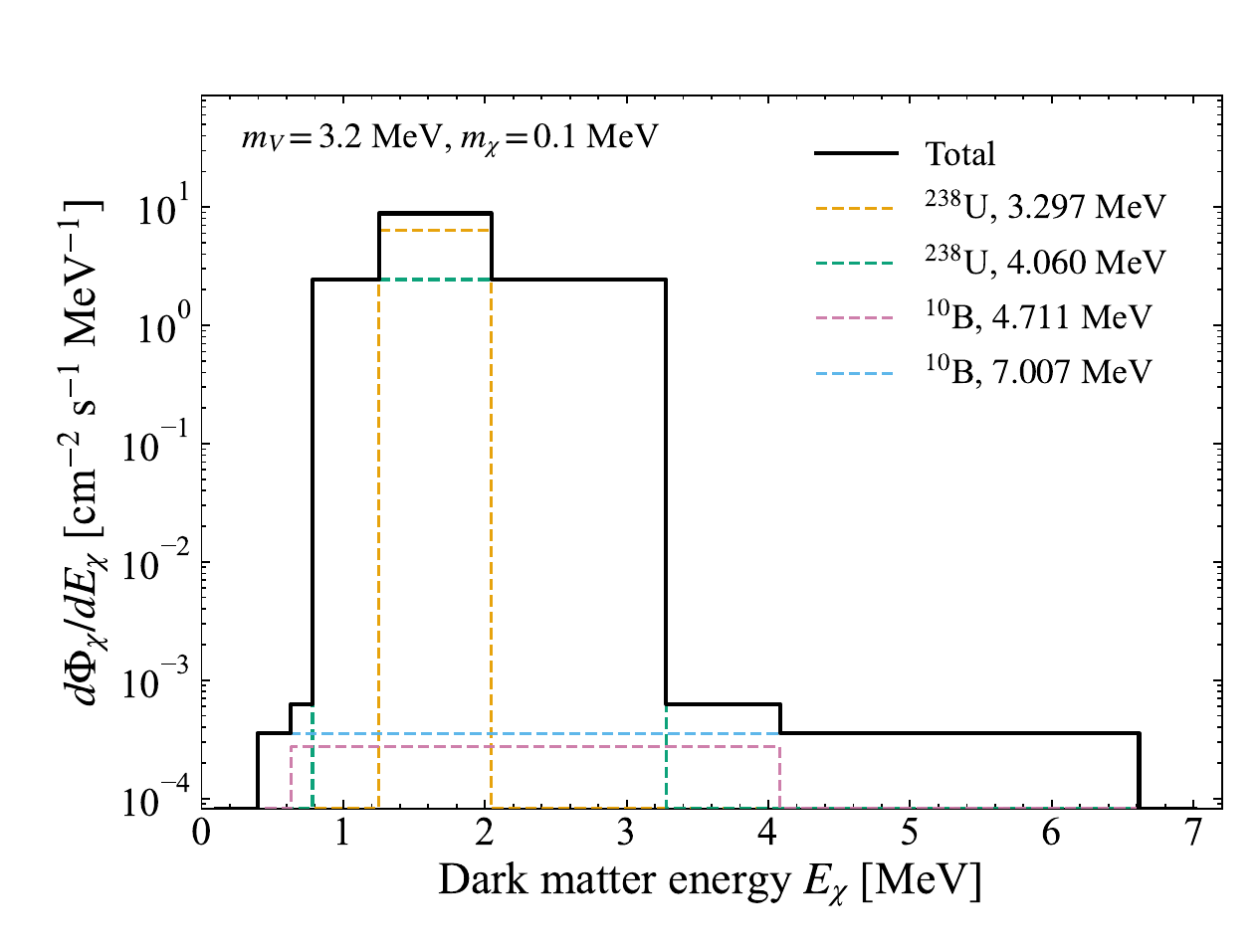}
\caption{Reactor dark matter flux from neutron-capture nuclear
de-excitation for the benchmark parameters
$(m_V,m_\chi)=(3.2,0.1)~{\rm MeV}$, $\epsilon=10^{-5}$, and
$\alpha_D=0.1$. Each transition produces a box-shaped spectrum from
$V\to\chi\bar{\chi}$, and the total flux is the sum over the retained
lines.}
\label{fig:dm_flux}
\end{figure}

We use a benchmark set of intense E1 neutron-capture transitions: the
\(3.297\) and \(4.060~{\rm MeV}\) lines from
\(^{238}{\rm U}(n,\gamma){}^{239}{\rm U}\), and the \(4.711\) and
\(7.007~{\rm MeV}\) lines from
\(^{10}{\rm B}(n,\gamma){}^{11}{\rm B}\). The present source model
includes only mediators produced directly in these primary nuclear
de-excitation transitions. Secondary production via
\(\gamma e^-\to Ve^-\), as well as contributions from additional
capture isotopes or other multipolarities, are not included. A more
complete nuclear de-excitation database could modify both the
normalization and the spectral shape of the predicted flux. The
resulting benchmark dark matter flux is shown in
Fig.~\ref{fig:dm_flux}.

\section{Migdal Signal and TEXONO Limits}
\label{sec:detection}

We compute the response of a low-threshold germanium detector to the
reactor-produced dark matter flux derived in
Sec.~\ref{sec:production}. The response is built from elastic nuclear
scattering, nuclear recoil quenching, and the single-ionization Migdal
contribution. For a single-ionization Migdal process, the total deposited energy is
\begin{equation}
E_{\rm det}
=
q_{\rm nr}(E_{\rm nr})E_{\rm nr}
+
E_{nl}
+
E_e,
\label{eq:Edet_migdal}
\end{equation}
where \(E_{\rm nr}\) is the nuclear recoil energy, \(E_{nl}\) is the
binding energy of the shell \((n,l)\), and \(E_e\) is the kinetic energy
of the emitted electron. 
Since the detector is sensitive to ionization, only the fraction of the nuclear-recoil energy transferred to electronic excitations contributes to the measured signal. This effect is
characterized by the nuclear-recoil quenching factor
\(q_{\rm nr}(E_{\rm nr})\), which gives the fraction of the recoil energy
that is visible in the detector.
We evaluate \(q_{\rm nr}\) using the Lindhard prescription~\cite{Lindhard:1963},
\begin{equation}
q_{\rm nr}(E_{\rm nr})
=
\frac{k\,g(\varepsilon)}{1+k\,g(\varepsilon)},
\qquad
g(\varepsilon)
=
3\varepsilon^{0.15}
+0.7\varepsilon^{0.6}
+\varepsilon,
\label{eq:lindhard}
\end{equation}
with $\varepsilon=11.5\,Z^{-7/3}(E_{\rm nr}/{\rm keV})$ and
$k=0.133\,Z^{2/3}A^{-1/2}$, which gives $k\simeq0.157$ for germanium.

The corresponding differential Migdal event rate is then obtained by combining the reactor dark matter flux, the nuclear scattering cross section, and the Migdal ionization probability,
\begin{equation}
\begin{split}
\frac{dR_{\rm Migdal}}{dE_{\rm det}}
&=
N_T
\int dE_\chi\,
\frac{d\Phi_\chi}{dE_\chi}
\int dE_{\rm nr}\,
\frac{d\sigma_{\chi A}}{dE_{\rm nr}}
\\
&\quad\times
\sum_{n,l}
\int dE_e\,
\frac{dp_v(nl\to E_e)}{dE_e}
\\
&\quad\times
\delta\!\left(
E_{\rm det}
-q_{\rm nr}(E_{\rm nr})E_{\rm nr}
-E_{nl}
-E_e
\right).
\end{split}
\label{eq:dRdEdet}
\end{equation}
Here \(N_T\) is the number of germanium target nuclei per unit detector mass, \(p_v\) denotes the Migdal ionization probability, and \(dp_v(nl\to E_e)/dE_e\) is the differential probability for ionizing an electron from the \((n,l)\) shell with kinetic energy \(E_e\). We compute the Migdal single-ionization probabilities in the isolated-atom approximation. For each occupied germanium orbital from the \(K\) through \(N\) shells (\(1s\), \(2s\), \(2p\), \(3s\), \(3p\), \(3d\), \(4s\), and \(4p\)), the bound and continuum electron wave functions are generated with cFAC version 1.7.1~\cite{Stambulchik:cFAC171}, a C-based fork of the Flexible Atomic Code~\cite{Gu:2008FAC}, and the ionization probability is evaluated from the overlap between the boosted initial electronic state and the final continuum state induced by the sudden nuclear recoil.

Previous studies of the Migdal effect in noble-element targets such as
xenon and argon have commonly described the electronic response using
atomic ionization probabilities~\cite{Ibe:2017yqa,Bell_2022},
while subsequent studies of semiconductor targets have explicitly
included the crystal environment~\cite{Essig:2019xkx,Knapen:2020aky}.
Here we use isolated-atom Migdal ionization probabilities for germanium
and assess this approximation by comparing the relevant atomic and
crystal scales. The germanium lattice constant
$a\simeq0.566~{\rm nm}$ corresponds to
$p_{\rm lat}\sim\hbar(2\pi/a)\simeq2.2~{\rm keV}/c$, whereas the
$M$-shell binding energies $E_B\simeq35$--$170~{\rm eV}$ imply
$p_{\rm atom}\sim\sqrt{2m_eE_B}\simeq6$--$13~{\rm keV}/c$.
With the germanium band gap $E_g\simeq0.67~{\rm eV}$~\cite{Ekuma_2013},
these binding energies are also about $50$--$250$ times larger than
$E_g$. These scale comparisons support the isolated-atom treatment for
the inner-shell contribution considered here. To remain conservative,
we impose $v_N/c\ge10^{-4}$.

The nuclear scattering cross section entering Eq.~\eqref{eq:dRdEdet}
is normalized by the reference DM--proton cross section
\begin{equation}
\bar{\sigma}_{\chi p}
=
\frac{16\pi\alpha\alpha_D\epsilon^2\mu_{\chi p}^2}
{(q_0^2+m_V^2)^2},
\qquad
q_0=\alpha m_e.
\label{eq:sigbar}
\end{equation}
For a germanium target with $Z=32$ and nuclear mass $m_A$, we use
\begin{equation}
\frac{d\sigma_{\chi A}}{dE_{\rm nr}}
=
\frac{
Z^2\bar{\sigma}_{\chi p}
F^2(E_{\rm nr})
|F_{\rm DM}(q)|^2
}{
4\mu_{\chi A}^2(E_\chi^2-m_\chi^2)
}
\mathcal{K}(E_{\rm nr}),
\label{eq:dsigdEnr}
\end{equation}
with the Helm form factor $F(E_{\rm nr})$~\cite{Helm:1956zz},
$q=\sqrt{2m_AE_{\rm nr}}$, the mediator form factor $|F_{\rm
DM}(q)|^2=(q_0^2+m_V^2)^2/(q^2+m_V^2)^2$, and the kinematic factor
$\mathcal{K}(E_{\rm nr})=2m_AE_\chi^2-E_{\rm nr}(2m_AE_\chi+m_\chi^2
+m_A^2-m_AE_{\rm nr})$. The Migdal probability is multiplied as an
atomic transition factor and does not introduce an additional nuclear
form factor.

\begin{figure}[!t]
\centering
\includegraphics[width=\columnwidth]{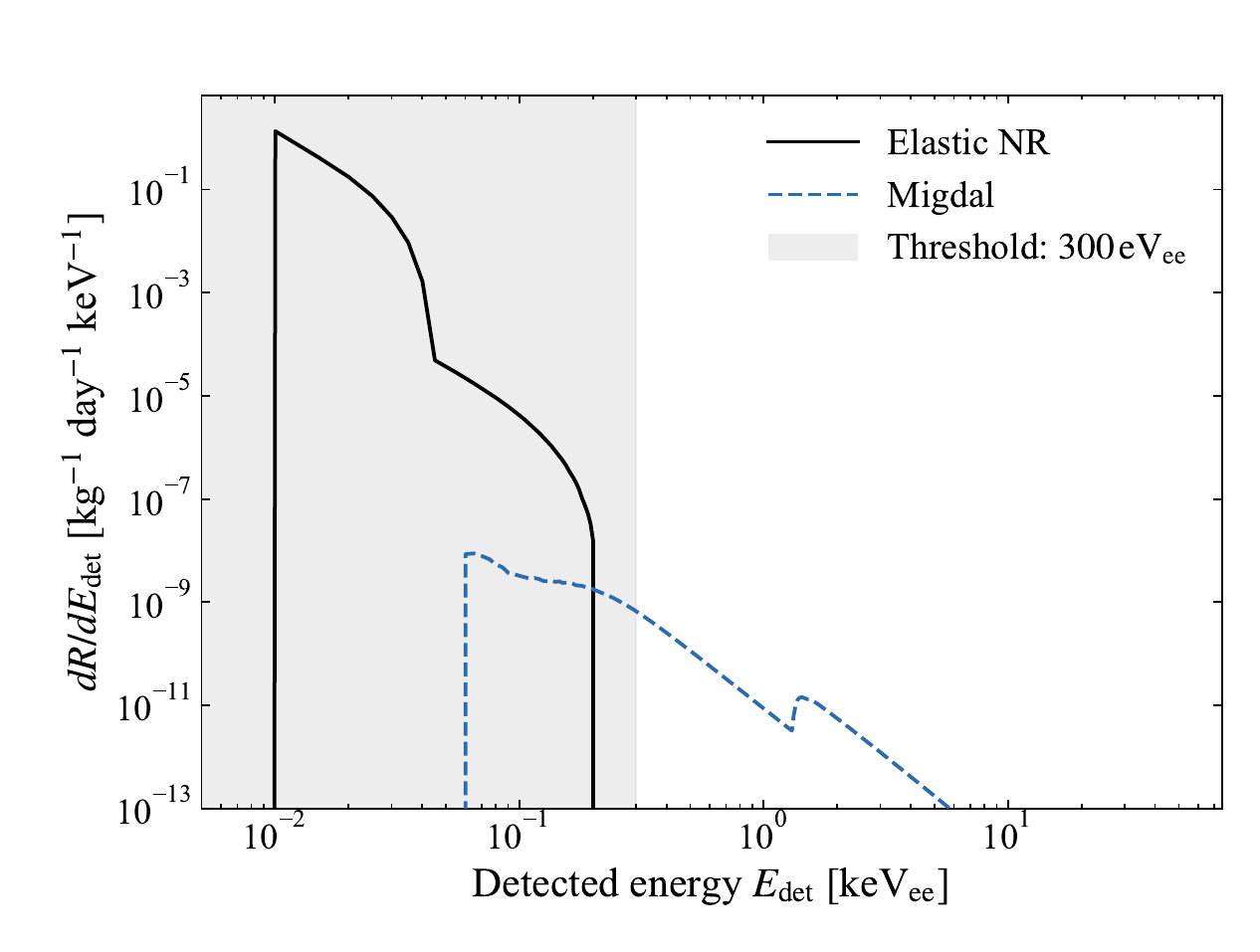}
\caption{Differential event rate of reactor-produced light dark
matter in the TEXONO germanium detector. The quenched elastic
nuclear-recoil contribution lies below the $300~{\rm eV_{ee}}$
analysis threshold for the benchmark shown, while Migdal ionization
populates the observable electron-equivalent energy region. The resulting
shell-resolved tables for \(dP_v(nl\to E_e)/dE_e\) are used in
Eq.~\eqref{eq:dRdEdet}, retaining only kinematically allowed
single-ionization events; the successive onset of each orbital's
threshold, \(E_{\rm det}\gtrsim E_{nl}+q_{\rm nr}E_{\rm nr}\), produces
the characteristic changes of slope in the Migdal spectrum shown in
Fig.~\ref{fig:ge_response}. Since the quenched elastic recoil lies
below threshold for the reactor-produced MeV-scale flux, the
observable signal is dominated by Migdal ionization.}
\label{fig:ge_response}
\end{figure}

For the benchmark \((m_V,m_\chi)=(3.2,0.1)\,{\rm MeV}\), the highest retained de-excitation line, \(7.007~{\rm MeV}\), gives \(E_\chi^{\rm max}\simeq6.61~{\rm MeV}\) and hence \(E_{\rm nr}^{\rm max}\simeq1.29~{\rm keV_{nr}}\). After Lindhard quenching, the maximum detected energy from the elastic nuclear recoil is about \(0.23~{\rm keV_{ee}}\), below the TEXONO threshold. Thus, the elastic NR curve in Fig.~\ref{fig:ge_response} ends at the quenched-recoil kinematic endpoint. Migdal ionization adds the electronic energy \((E_{nl}+E_e)\) to the detected signal, allowing events to enter the observable window. For this benchmark, the above-threshold Migdal rate is dominated by the \(M\) and \(L\) shells, contributing about \(85.6\%\) and \(13.0\%\), respectively, while the \(N\) and \(K\) shells contribute only \(1.15\%\) and \(0.23\%\). The \(M\)-shell binding energies lie in the range \(\sim35\text{--}170~{\rm eV}\), and the prominent change of slope near \(E_{\rm det}\simeq1.2\text{--}1.4~{\rm keV}\) is associated with the \(L\)-shell contribution, dominated by the \(2p\) orbital. The \(K\)-shell threshold near \(11~{\rm keV}\) lies in the low-rate tail.

To constrain this channel, we derive constraints from the publicly available TEXONO reactor ON--OFF residual spectrum, defined as the difference between the spectra measured with the reactor operating and shut down, using data from the low-threshold germanium detector at the Kuo-Sheng Reactor Neutrino Laboratory~\cite{Singh_2019}. Since the dark matter flux is produced by the reactor, it is retained in this residual, while reactor-independent backgrounds are reduced by the subtraction. We therefore use the residual spectrum as the observable for the constraint analysis. For each mass point $(m_\chi,m_V)$, the bin-averaged signal is
\begin{equation}
S_b(\epsilon,m_V,m_\chi)
=
\frac{1}{\Delta E_b}
\int_{E_b^{\rm min}}^{E_b^{\rm max}}
dE_{\rm det}\,
\frac{dR_{\rm Migdal}}{dE_{\rm det}},
\label{eq:bin_signal}
\end{equation}
where $E_b^{\rm min}$ and $E_b^{\rm max}$ are the lower and upper edges
of the $b$-th detector-energy bin and
$\Delta E_b=E_b^{\rm max}-E_b^{\rm min}$. We define the Gaussian
chi-square as
\begin{equation}
\chi^2(\epsilon,m_V,m_\chi)
=
\sum_b
\frac{
\left[
R_b-S_b(\epsilon,m_V,m_\chi)
\right]^2
}{
\sigma_b^2
},
\label{eq:chi2}
\end{equation}
where $R_b$ and $\sigma_b$ are the measured ON--OFF residual rate and
its uncertainty in the $b$-th bin. We minimize $\chi^2$ over
$\epsilon\ge0$ and obtain the one-sided 95\% C.L. upper limit
$\epsilon_{\rm lim}$ from
$\Delta\chi^2(\epsilon_{\rm lim})=2.71$, with
$\Delta\chi^2(\epsilon)=\chi^2(\epsilon)-\chi^2_{\rm min}$. The mediator
mass scan covers $3.2~{\rm MeV}\le m_V\le6.9~{\rm MeV}$, with the lower
bound chosen to isolate the neutron-capture de-excitation source from
the lower-mass region where Compton-like reactor production can also
contribute~\cite{lou2026probingdarkphotonsnuclear}, and the upper bound
set below $\omega_{\rm max}=7.007~{\rm MeV}$, where the available
de-excitation phase space vanishes.

\begin{figure}[!t]
\centering
\includegraphics[width=\columnwidth]
{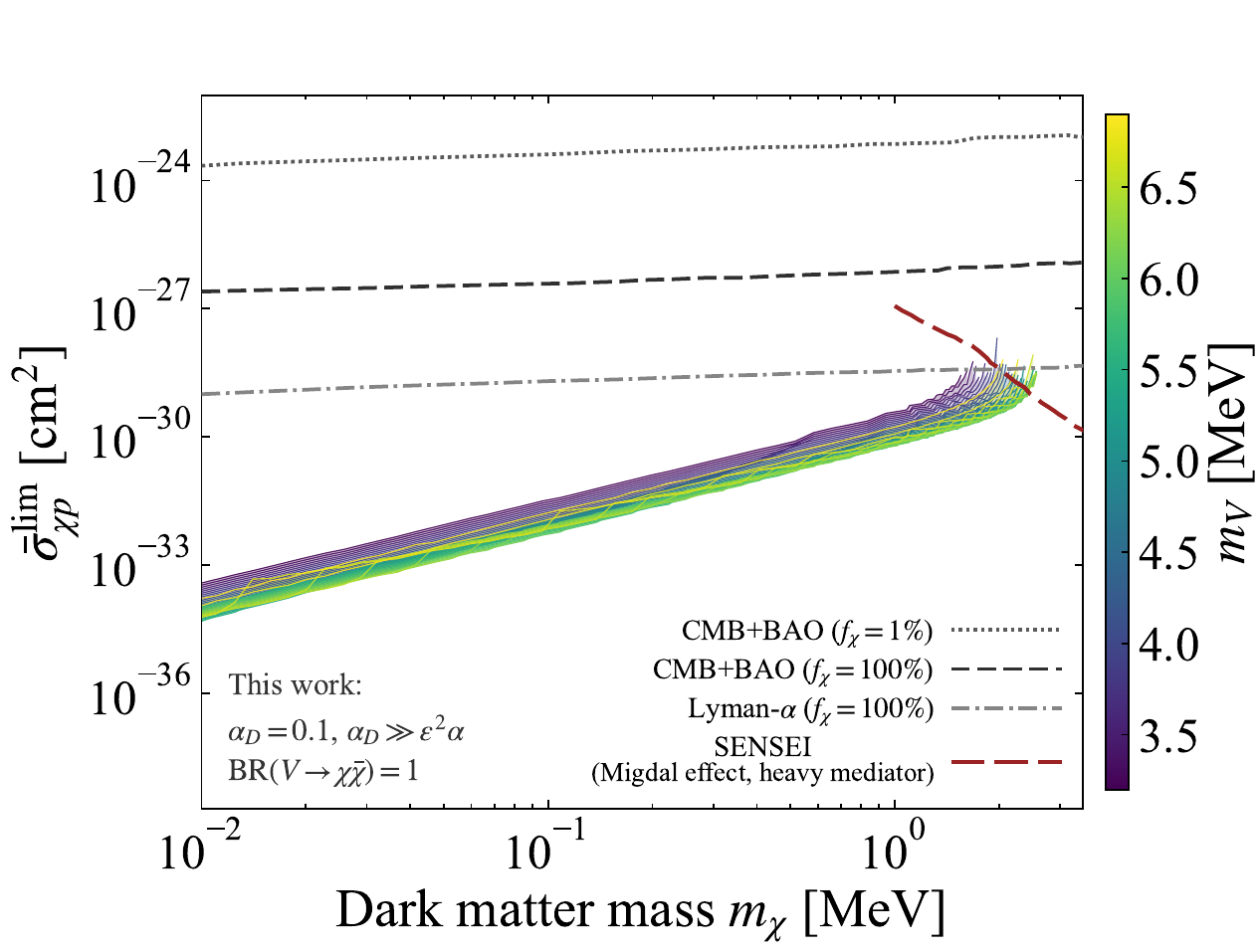}
\caption{TEXONO 95\% C.L. upper limits on $\bar{\sigma}^{\rm lim}_{\chi p}$ from reactor-produced light dark matter via Migdal ionization in the low-threshold germanium detector, shown as a function of $m_\chi$ for $m_V$ in the range $3.2$--$6.9~{\rm MeV}$ (color scale). The limits are obtained in the kinetically mixed dark photon scenario
with $\alpha_D=0.1$ and $\alpha_D\gg\epsilon^2\alpha$, taking the
invisible branching ratio to be
${\rm BR}(V\to\chi\bar{\chi})=1$.
Existing constraints are shown for comparison but are not derived
within the same theoretical framework. The CMB+BAO and Lyman-$\alpha$
95\% C.L. bounds constrain a velocity-independent
DM--proton interaction ($n=0$)~\cite{Buen-Abad:2021mvc}, with the assumed
interacting dark matter fraction $f_\chi$ indicated in the figure. The
SENSEI 90\% C.L. Migdal (MG) limit assumes Galactic halo dark matter
scattering from silicon nuclei through a heavy scalar mediator, with the
observable ionization signal generated by the Migdal effect
~\cite{SENSEI:2023zdf}. In contrast to the cosmological bounds, the
reactor limits depend on the laboratory-produced $\chi\bar{\chi}$ flux
and do not require $\chi$ to constitute any specified fraction of the
cosmological dark matter abundance.}
\label{fig:limits_dm}
\end{figure}

The resulting 95\% C.L. upper limits on $\bar{\sigma}_{\chi p}$ are
shown in Fig.~\ref{fig:limits_dm}. For fixed $m_\chi$, larger $m_V$
values are supported mainly by the higher-energy retained transitions,
which typically yield more energetic $\chi\bar{\chi}$ pairs. This
enhances the Migdal signal in the observable energy window and
strengthens the limits over most of the scan range. Close to the
endpoint of the highest retained transition, however, the production
phase space is suppressed, and the limits weaken as $m_V$ approaches
$\omega_{\rm max}$.  Theoretically, the mass of dark matter can reach
\(m_\chi<\omega_{\rm max}/2\simeq3.50~{\rm MeV}\).
However, the present analysis loses sensitivity for
\(m_\chi\gtrsim2.56~{\rm MeV}\) because of the negligible event rate. As \(m_\chi\) approaches this upper end of the scan, the
available phase space and the above-threshold event rate decrease
rapidly, producing the upward slope of the limit curves.

For comparison, we show the CMB+BAO and Lyman-$\alpha$ constraints of
Ref.~\cite{Buen-Abad:2021mvc}, together with the SENSEI direct detection
limit~\cite{SENSEI:2023zdf}. These existing bounds are not derived within
the same theoretical framework as our reactor constraint. The CMB+BAO
and Lyman-$\alpha$ bounds are obtained for a velocity-independent
DM--proton interaction ($n=0$), with the assumed interacting dark matter
fraction $f_\chi$ indicated in Fig.~3. The SENSEI limit assumes Galactic
halo dark matter scattering from silicon nuclei through a heavy scalar
mediator, with the observable ionization signal generated by the Migdal
effect. By contrast, our reactor limits are obtained in the kinetically
mixed dark photon scenario and depend on the laboratory-produced
$\chi\bar{\chi}$ flux rather than on the Galactic halo or cosmological
abundance of $\chi$.
Relevant cosmological constraints also include BBN and
\(N_{\rm eff}\) bounds, which depend on the thermalization,
decoupling, and entropy-transfer history of the dark
sector~\cite{Giovanetti:2021izc,Ge:2024cto}, while freeze-in targets
depend on both the production history and the relic abundance assigned
to \(\chi\)~\cite{Chang:2019xva}.  Astrophysical
constraints from stellar and supernova cooling depend on the
in-medium production, absorption, trapping, and transport of
dark-sector states~\cite{Chang:2019xva,Chang:2018rso}. In particular,
when dark self-interactions are strong, the produced states may form a
coupled dark fluid rather than free-streaming particles, which can
substantially modify the resulting constraints~\cite{Fiorillo:2024upk}.

\section{Conclusion}
\label{sec:conclusion}

We have studied Migdal ionization as an observable channel for
reactor-produced sub-MeV dark matter in low-threshold germanium
detectors. The reactor source is provided by neutron-capture nuclear
de-excitation, in which excited nuclei can emit on-shell kinetically
mixed vector mediators in place of ordinary photons. The mediators
subsequently decay invisibly into $\chi\bar{\chi}$, producing dark
matter with box-shaped energy spectra.

Using selected E1 transitions from $^{238}{\rm
U}(n,\gamma){}^{239}{\rm U}$ and $^{10}{\rm B}(n,\gamma){}^{11}{\rm
B}$, we computed the resulting reactor dark matter flux at the TEXONO
baseline, and calculated the germanium detector response including
elastic nuclear scattering, nuclear recoil quenching, and Migdal
ionization. For the benchmark spectra considered here, the nuclear recoil contribution alone remains below the $300~{\rm
eV_{ee}}$ analysis threshold, while Migdal ionization adds electronic energy to
the detector response and provides the relevant visible channel. From the TEXONO reactor ON--OFF residual spectrum, we derived 95\%
C.L. upper limits on the reference DM--proton cross section
$\bar{\sigma}_{\chi p}$, ranging roughly from $6.7\times10^{-35}$ to
$8.0\times10^{-33}~{\rm cm^2}$, for MeV-scale mediators and dark
matter masses $0.01~{\rm MeV}<m_\chi<2.6~{\rm MeV}$. These limits rely only on the
reactor-produced $\chi\bar{\chi}$ flux and the germanium detector
response, and do not assume that $\chi$ constitutes the cosmological
dark matter abundance.


\begin{acknowledgments}
This work is supported by 
the NNSFC under Grant No.~12275134 and No.~12335005.
\end{acknowledgments}

\bibliographystyle{apsrev4-2}
\bibliography{refs}

\end{document}